\documentclass{PoS}
\def\mpi2{m_\pi^2}
\def\mK2{m_K^2}

\newcommand{\bea}{\begin{eqnarray}}
\newcommand{\eea}{\end{eqnarray}}
\newcommand{\be}{\begin{equation}}
\newcommand{\ee}{\end{equation}}

\newsavebox{\DERIVBOXZLM}
\savebox{\DERIVBOXZLM}[2.5em]{$\Longrightarrow\hspace{-1.5em}
\raisebox{.2ex}{*}
\hspace{-.7em}\raisebox{-.8ex}{\scriptsize lm}\hspace{.7em}$}

\title{Effects of the charm quark on the QCD equation of state}

\ShortTitle{Effects of the charm quark on the QCD equation of state}

\author{\speaker{Ludmila Levkova}%
         \thanks{MILC Collaboration}\\
        Physics Department, University of Utah, Salt Lake City, UT 84112, USA\\
        E-mail: \email{ludmila@physics.utah.edu}}

\abstract{We study the effects of the addition of the charm quark on the QCD equation of state at zero and nonzero chemical
potential on lattices with $N_t=6$.
Our ensembles are quenched with respect to charm and the charm quark
is a valence staggered quark. Along the trajectory of constant
physics the ratio $m_s/m_c$ is kept constant after tuning the
charm quark mass at a lattice spacing of about 0.09 fm. We find that
the charm quark has a significant contribution to the
equation of state at zero chemical potential already at temperatures between about $1.2T_c$ and $2T_c$. 
The additional contribution
at nonzero chemical potential vanishes within the current statistical uncertainty.
          }

\FullConference{The XXVII International Symposium on Lattice Field Theory\\
                 July 26-31, 2009\\
                 Peking University, Beijing, China}

\begin{document}

\section{Introduction}
The equation of state (EOS) of the quark-gluon plasma is an essential input to the hydrodynamics 
models which are used to interpret the experimental data from heavy-ion collisions.
The experiments at RHIC create a fire ball which thermalizes within $\tau\approx 10^{-24}$ s \cite{adcox}.
The $u$, $d$ and $s$ quarks participate in the thermal ensemble describing the state of the thermalized
fire ball. On the other hand, under the experimental conditions the $c$ quark probably
is not thermalized and thus  the 2+1 flavor EOS is considered sufficient
for the hydrodynamics models applied to the experimental data.
However, a quark-gluon plasma existed microseconds after the Big Bang. Under these primordial conditions
the $c$ quark probably participated in the thermal ensemble as well, which implies that for 
the study of the early Universe, the EOS with 2+1+1 flavors might be important.
Previously, the question of the charmed quark contribution to the EOS at zero chemical potential
has been studied in \cite{mcheng}. That work was performed with the p4 action on lattices with $N_t=4$.

In the present study, we determine the charm quark contribution to the EOS at zero and nonzero
chemical potential. The calculations are done on the 
MILC collaboration $2+1$ flavor asqtad lattices with $m_l=0.1m_s$, where $m_s$
is tuned close to the
strange quark mass. The nonzero 
temperature calculations are done on lattices with $N_t=6$.
We simulate on trajectories of constant physics.
Similarly to Ref.~\cite{mcheng}, the charm quark is just a valence staggered quark.
The charm quark mass is tuned to match the rest mass of the $D_s$ at $\beta=7.08$ 
($a\approx0.086$~fm)
where the discretization effects are smallest on our trajectory. Within 4\%
 at that point, $m_c/m_s=10$.
We have kept that ratio constant for lower temperatures. The tuning is probably 
incorrect at the lowest 
temperatures, but we do not expect this to matter much due to the large mass of the $c$ quark and 
its small contribution there.
The nonzero chemical potential calculation is done using the Taylor expansion method 
\cite{tm} to sixth order. We used 800 random sources
in the transition region and 400 outside it.
For the calculation at nonzero chemical potential, the valence $c$ quark has a low cost 
in terms of computer time, but it requires
a sizable software development. 
For $2+1$ flavors we have $95$ observables to code and for  $2+1+1$
flavors there are $399$.

We determine the EOS at zero chemical potential using the integral method \cite{im,mine},
in which the pressure is obtained from the interaction measure (trace anomaly)
by integrating it along a trajectory of constant physics.  
At nonzero chemical potential, the Taylor expansion method gives for the pressure
\be
{p\over T^4}=
\sum_{n,m,k=0}^\infty c_{nmk}(T) \left({\bar{\mu}_l\over T}\right)^n
\left({\bar{\mu}_h\over T}\right)^m\left({\bar{\mu}_c\over T}\right)^k,
\ee
where $\bar{\mu}_{l,h,c}$ are the chemical potentials in physical units for the light ($u,d$), heavy ($s$)
and charm ($c$) quarks.
Due to CP symmetry the terms in the above are nonzero only if $n+m+k$ is even.
The coefficients are
\be
c_{nmk} (T)=
{1\over n!}{1\over m!}{1\over k!}{N_t^{3}\over N_\sigma^3}{{\partial^{n+m+k}
\ln{\cal Z}}\over{\partial(\mu_l N_t)^n}{\partial(\mu_h N_t)^m}
{\partial(\mu_c N_t)^k}}\biggr\vert_{\mu_{l,h,c}=0}.
\ee 
A similar expression can be written for the interaction measure coefficients.

\section{The EOS at zero chemical potential}
\begin{figure}[ht]
\begin{center}
\includegraphics[width=0.8\textwidth]{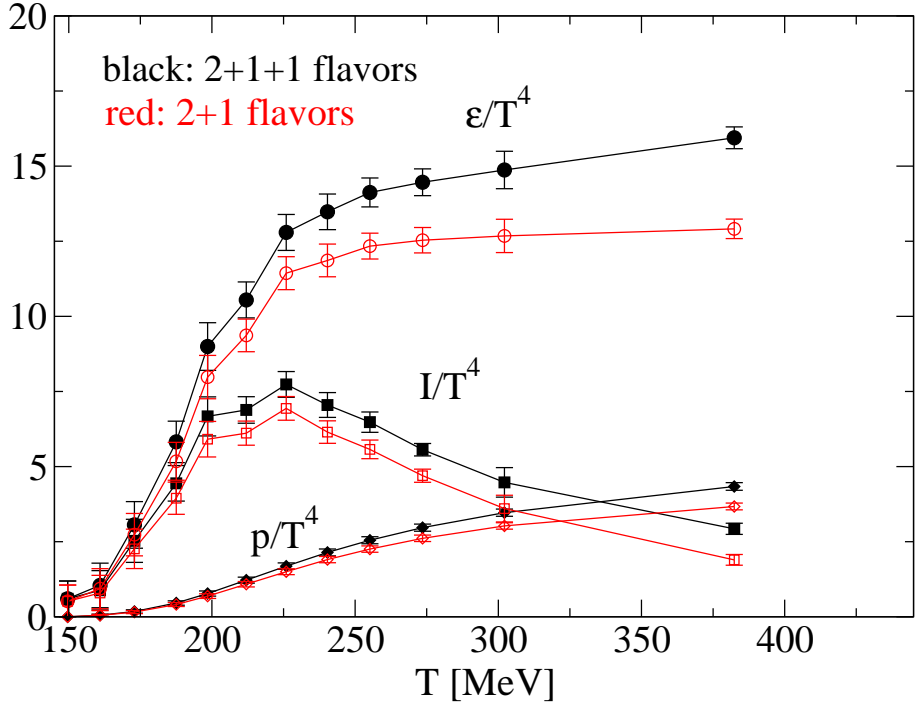}
\caption{Interaction measure ($I$), pressure ($p$) and energy density ($\varepsilon$)
divided by the temperature to the fourth power ($T^4$) for the cases of 2+1 and 2+1+1
flavors.
}
\label{fig:zeromuEOS}
\end{center}
\end{figure}
Figure~\ref{fig:zeromuEOS} shows our results for the EOS with 2+1+1 flavors
and compares it with the previously determined 2+1 flavor case \cite{mine}. The charm 
quark contribution grows with temperature, as expected and at the highest
available $T$ it contributes about 20\% to the energy density. We conclude that
in the cases where the charm quark is thermalized, its contribution  
to the EOS at temperatures
higher than about 200 MeV, cannot be simply ignored. Our result is qualitatively similar to the
previous work in Ref.~\cite{mcheng}.

\section{The EOS at nonzero chemical potential}
\begin{figure}[ht]
\begin{center}
\includegraphics[width=0.8\textwidth]{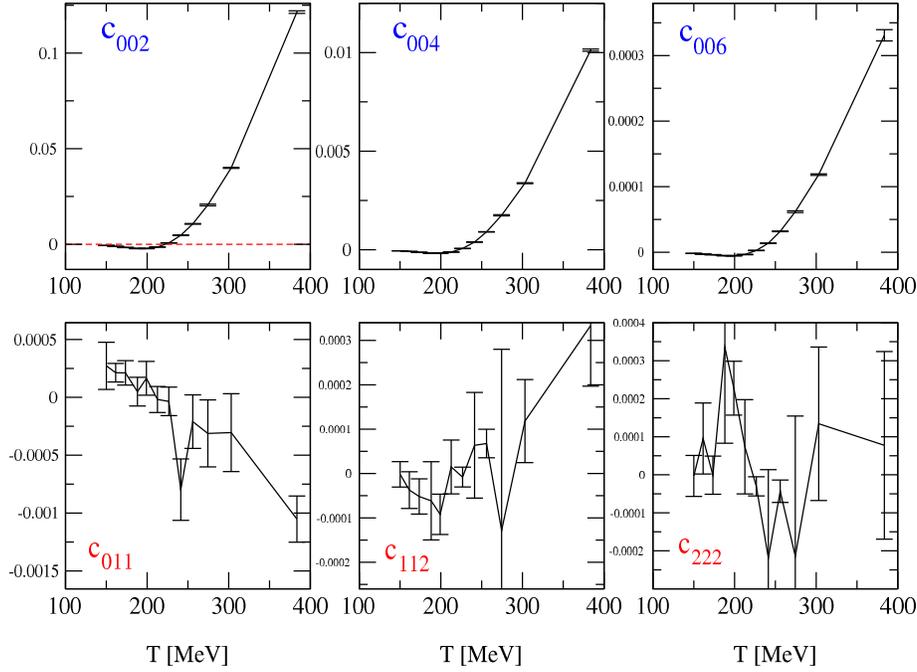}
\caption{Some of the expansion coefficients for the pressure at nonzero chemical potential.} 
\label{fig:Pcharm_coeff}
\end{center}
\end{figure}
Figure~\ref{fig:Pcharm_coeff} presents our results for some of the pressure
expansion coefficients which are directly related to the charm quark contribution
at nonzero chemical potential. The first row shows the "unmixed" coefficients
and the second row - three of the "mixed" coefficients (they involve derivatives with respect to
the light and heavy chemical potentials in addition to the charm one). The mixed 
coefficients are much noisier than the unmixed ones. We also find an unphysical behavior
for the $c_{002}$ coefficient, which around the transition region turns slightly negative.
Possible causes could be the quenched approximation for the charm quark
and the large heavy quark discretization effects. We intend to study this behavior in the near future.
We also have results for the coefficients in the interaction measure expansion, 
but due to lack of space, we do not include them here. They will be published in a future
work.

\begin{figure}[ht]
\begin{center}
\includegraphics[width=0.8\textwidth]{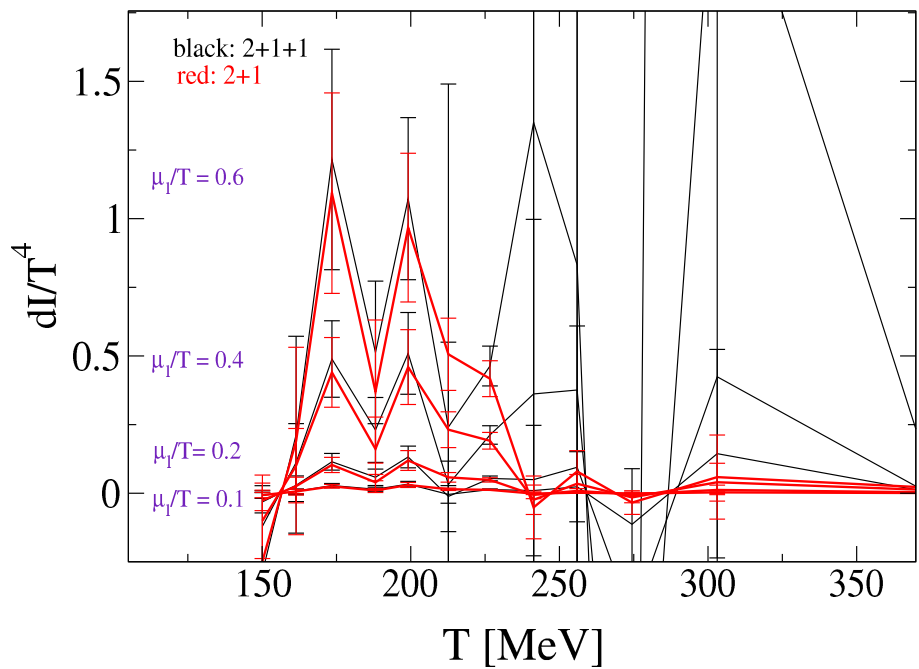}
\caption{The correction to the interaction measure due to the nonzero chemical potentials
(with the strange and charm chemical potentials restricted to values which give zero
strange and charm quark densities).} 
\label{fig:dIcharm}
\end{center}
\end{figure}

\begin{figure}[ht]
\begin{center}
\includegraphics[width=0.85\textwidth]{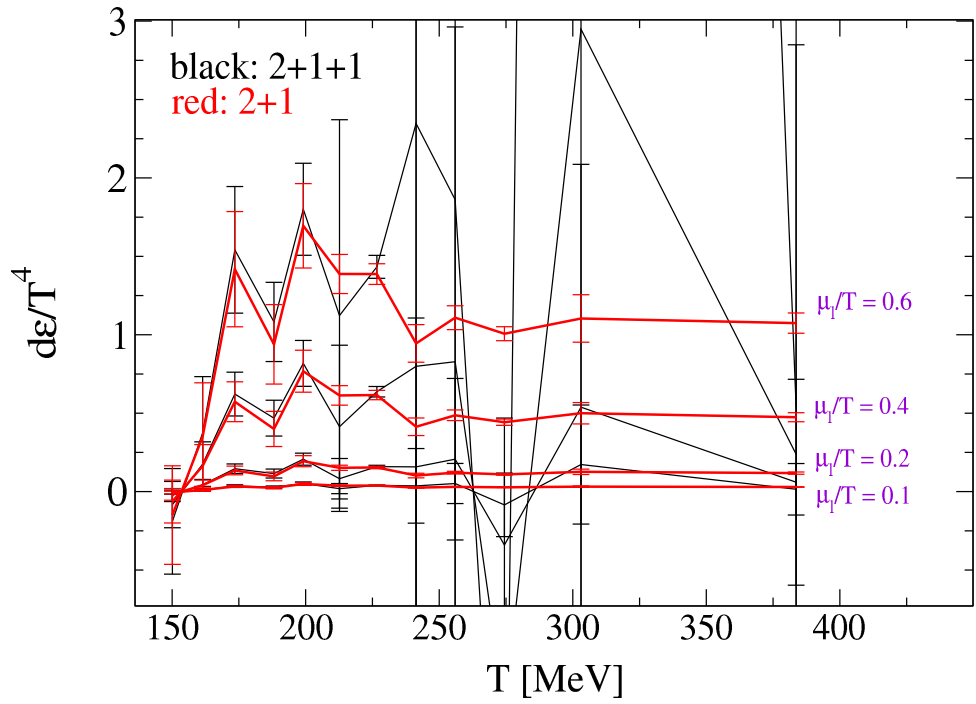}
\caption{The correction to the energy density due to the nonzero chemical potentials
}
\label{fig:dEcharm}
\end{center}
\end{figure}

The results for the contribution of the chemical potentials to the interaction measure
and the energy density for the 2+1+1 flavor case are presented in Figs.~\ref{fig:dIcharm}~and~\ref{fig:dEcharm},
respectively. The results are for different values of $\mu_l/T$, while
$\mu_{h,c}$ are tuned such that the heavy (strange) and the charm quark
densities are zero within statistical errors. In these figures we also include our previous 2+1 flavor
results \cite{mine1}. The comparison between these two cases shows a negligible contribution due 
to the charm quark addition at low temperatures. At high temperatures the statistical errors 
are too large for a definite conclusion about the charm quark effects. 

\section{The isentropic EOS}
\begin{figure}[ht]
\begin{center}
\includegraphics[width=0.85\textwidth]{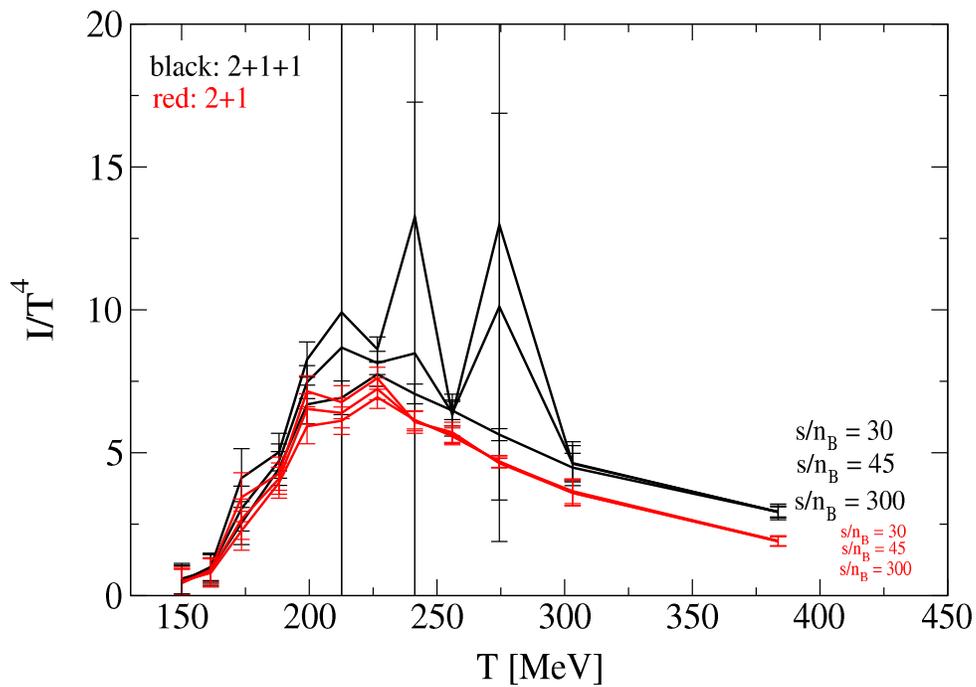}
\caption{The isentropic interaction measure normalized to the fourth power of temperature. 
}
\label{fig:Iisentr}
\end{center}
\end{figure}

\begin{figure}[ht]
\begin{center}
\includegraphics[width=0.85\textwidth]{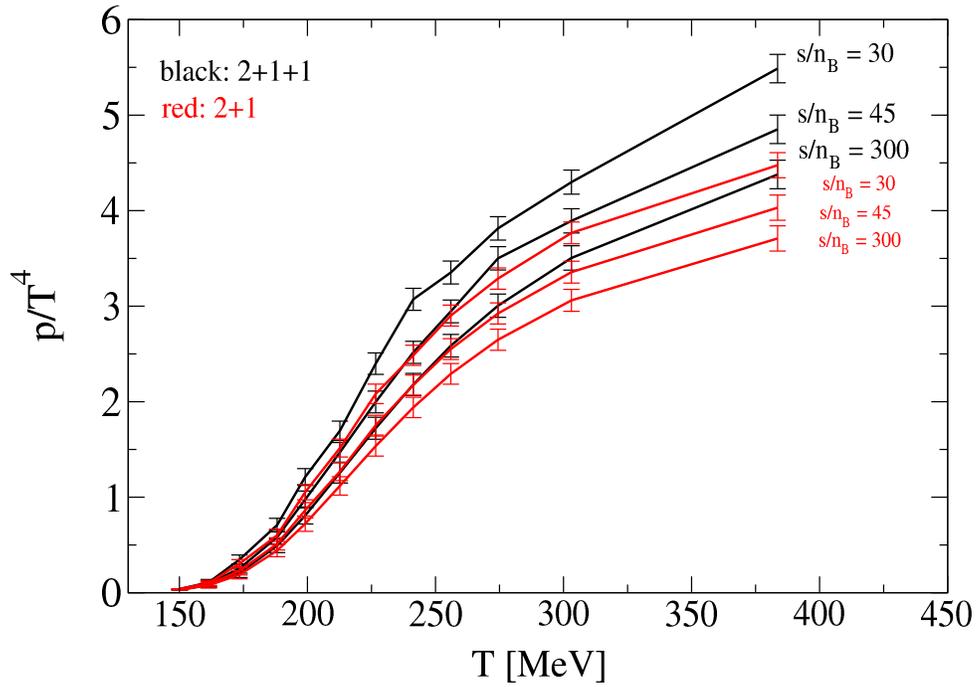}
\caption{The isentropic pressure normalized to the fourth power of temperature. 
}
\label{fig:Pisentr}
\end{center}
\end{figure}

\begin{figure}[ht]
\begin{center}
\includegraphics[width=0.85\textwidth]{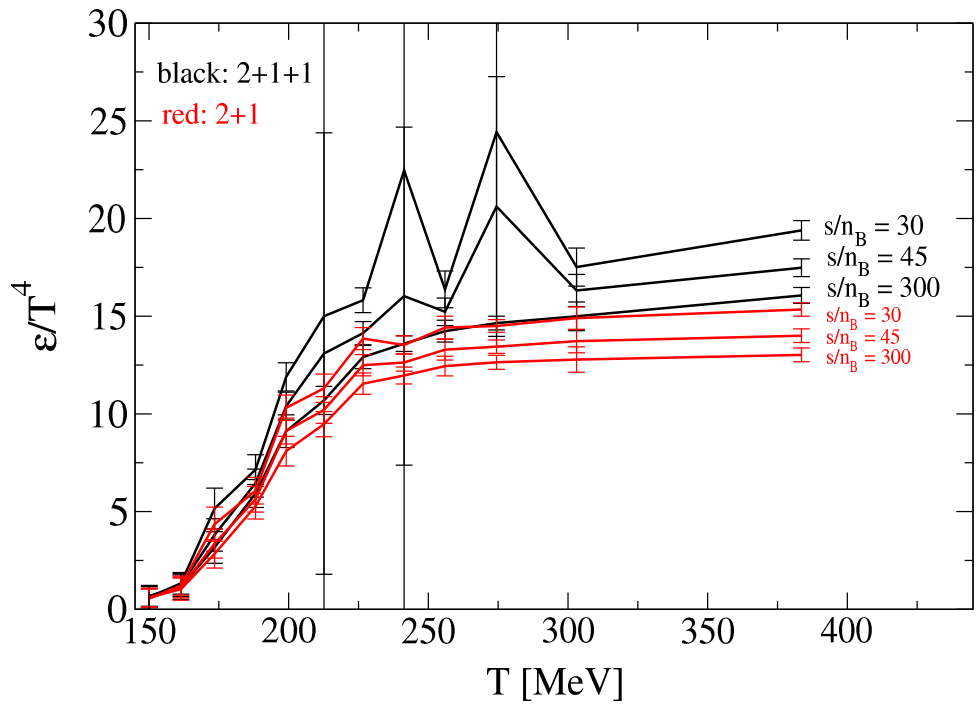}
\caption{  The isentropic energy density normalized to the fourth power of temperature.}
\label{fig:Eisentr}
\end{center}
\end{figure}

The heavy-ion collision experiments produce matter that expands isentropically. 
This implies that
$s/n_B$ remains constant. For the data obtained at the AGS, SPS, and RHIC experiments, $s/n_B$
is approximately 30, 45 and 300, respectively.
We have determined the approximate isentropic trajectories in 
the ($\mu_l$, $\mu_h$, $\mu_c$, $T$)
space, 
by numerically solving the system
\be
{s\over n_B}(\mu_l, \mu_h, \mu_c)= C,\hspace{0.5cm}
{n_s\over T^3} (\mu_l, \mu_h, \mu_c) = 0,\hspace{0.5cm}
{n_c\over T^3}(\mu_l, \mu_h, \mu_c) = 0,
\ee
with $C=30$, 45 and 300. The results for the interaction measure, pressure
and the energy density along these isentropic trajectories are shown in
Figs.~\ref{fig:Iisentr},~\ref{fig:Pisentr}~and~~\ref{fig:Eisentr}, for both the 2+1+1 and
the 2+1 flavor cases \cite{mine1}. The comparison indicates that the charm quark contribution is non-negligible,
although it is due mostly to the contribution of the zeroth order coefficients in the 
Taylor expansions ({\it i.e.}, the EOS calculated at zero chemical potential).
\section{Conclusions}
The addition of the charm quark has a noticeable effects 
on the EOS at zero chemical potential at high temperatures. 
This implies that for the EOS of the early Universe we cannot ignore
the charm quark contributions. It is possible that the rest of the flavors 
should be included too if high enough temperatures are studied.
The effects of the charm quark  on $dI$ and $d\varepsilon$ at nonzero chemical potential 
are small. Since the charm quark mass is large, many of the Taylor expansion coefficients which involve
derivatives with respect to $\mu_c/T$ 
do not reach the massless Stefan-Boltzmann values in the temperature range studied.
We also have found a negative "dip"
for $c_{002}$ around  the  transition temperature --- an unphysical effect whose origin we intend to
study in the near future.
And lastly, we found that the isentropic EOS is also affected by the addition of the charm
quark (due mainly to its large effect on the EOS at zero chemical potential.)

\end{document}